\documentclass[twocolumn,prl,superscriptaddress]{revtex4}

\usepackage{dcolumn}
\usepackage{amsmath}
\usepackage{graphicx}
\usepackage{subfigure}

\newcommand{\half}{\tfrac12}

\newcommand{\av}[1]{\left\langle#1\right\rangle}
\newcommand{\etal}{{\it{}et~al.}}

\newcommand{\Tr}{\mathop\mathrm{Tr}}
\newcommand{\erf}{\mathop\mathrm{erf}}
\renewcommand{\Im}{\mathop\mathrm{Im}}
\newcommand{\mat}{\mathbf}
\renewcommand{\vec}{\mathbf}

\newcommand\pin{p_\textrm{in}}
\newcommand\pout{p_\textrm{out}}
\newcommand\cin{c_\textrm{in}}
\newcommand\cout{c_\textrm{out}}

\begin{document}

\title{Graph spectra and the detectability of community structure in networks}
\author{Raj Rao Nadakuditi}
\affiliation{Department of Electrical Engineering and Computer Science,
  University of Michigan, Ann Arbor, MI 48109}
\author{M. E. J. Newman}
\affiliation{Department of Physics and Center for the Study of Complex
  Systems, University of Michigan, Ann Arbor, MI 48109}

\begin{abstract}
  We study networks that display community structure---groups of nodes
  within which connections are unusually dense.  Using methods from random
  matrix theory, we calculate the spectra of such networks in the limit of
  large size, and hence demonstrate the presence of a phase transition in
  matrix methods for community detection, such as the popular modularity
  maximization method.  The transition separates a regime in which such
  methods successfully detect the community structure from one in which the
  structure is present but is not detected.  By comparing these results
  with recent analyses of maximum-likelihood methods we are able to show
  that spectral modularity maximization is an optimal detection method in
  the sense that no other method will succeed in the regime where the
  modularity method fails.
\end{abstract}

\maketitle

The problem of community detection in networks has attracted a substantial
amount of attention in recent years~\cite{GN02,Fortunato10}.  Communities
in this context are groups of vertices within a network that have a high
density of within-group connections but a lower density of between-group
connections.  The challenge is to find such groups accurately and
efficiently in a given network---the ability to do so would have
applications in the analysis of observational data, network visualization,
and complexity reduction and parallelization of network problems.

In this paper we focus on matrix methods for community detection, which are
based on the properties of matrix representations of networks such as the
adjacency matrix or the modularity matrix.  While significant effort has
been devoted to the development of practical algorithms using these
methods, there has been less work on formal examination of their properties
and implications for algorithm performance.  Here we give an analysis of
the spectral properties of the adjacency and modularity matrices using
random matrix methods, and in the process uncover a number of results of
practical importance.  Chief among these is the presence of a sharp
transition between a regime in which the spectrum contains clear evidence
of community structure and a regime in which it contains none.  In the
former regime, community detection is possible and current algorithms
should perform well; in the latter, any method relying on the spectrum to
perform structure detection must fail.  A similar phase transition has been
reported recently in an analysis of a different class of detection methods,
based on Bayesian inference~\cite{DKMZ11a}.  By comparing the two analyses,
we are able to demonstrate that methods such as modularity maximization are
optimal, in the sense that no other method will succeed where they fail.

For the formal analysis of community structured networks, we must define
the particular network or networks we will study.  In this paper we focus
on the most widely studied model of community structure, the stochastic
block model, although our methods could be applied to other models as well.
The stochastic block model, in its simplest form, divides a network of $n$
vertices into some number~$q$ of groups denoted by $r=1\ldots q$ and then
places undirected edges between vertex pairs~$i,j$ with independent
probabilities~$p_{rs}$, where $r,s$ are respectively the groups to which
vertices~$i,j$ belong.  In other words, the probability of an edge between
two vertices in this model depends only on the groups in which the vertices
fall.  If the diagonal elements of the matrix of probabilities~$p_{rs}$ are
greater than the off-diagonal elements, then the network displays classic
community structure with a greater density of edges within groups than
between them.  Particular instances of the stochastic block model are
commonly used as testbeds for assessing the performance of community
detection algorithms---especially in the ``four groups'' test~\cite{GN02}
and the planted partition model~\cite{CK01}.

Let us first demonstrate our argument for the simplest possible case of a
network with $q=2$ groups of equal size~$\half n$ each and just two
different probabilities~$\pin$ and~$\pout$ for connections within and
between groups.  We focus particularly on the case of sparse networks,
those for which the fraction of possible edges that are present in the
network vanishes in the limit of large~$n$, which appears to be
representative of most networks observed in the real world, although our
results apply in principle to dense networks as well.

The adjacency matrix~$\mat{A}$ of an undirected network is the $n\times n$
symmetric matrix with elements $A_{ij}=1$ if vertices $i$ and~$j$ are
connected by an edge and 0 otherwise.  If we average the adjacency matrix
over the ensemble of our stochastic block model the resulting
matrix~$\av{\mat{A}}$ has elements equal to $\pin$ for vertices in the same
group and~$\pout$ for vertices in different groups.  Defining $\cin =
n\pin$ and $\cout = n\pout$, this matrix can be written in the form
\begin{equation}
\av{\mat{A}} = \half (\cin+\cout)\,\vec{1}\vec{1}^T
               + \half (\cin-\cout)\,\vec{u}\vec{u}^T,
\end{equation}
where $\vec{1}$ and $\vec{u}$ are the unit vectors $\vec{1} =
(1,1,1,\ldots)/\sqrt{n}$ and $\vec{u} =
(1,1,\ldots,-1,-1,\ldots)/\sqrt{n}$, the $\pm1$ elements in the latter
denoting the members of the two communities.

Now the full adjacency matrix can be written in the form $\mat{A} =
\av{\mat{A}} + \mat{X}$ where the matrix~$\mat{X}$ is the deviation between
the adjacency matrix and its average value.  By definition, $\mat{X}$~is a
symmetric random matrix with independent elements of mean zero.

Our analysis will focus on the spectrum of eigenvalues~$z$ of the adjacency
matrix, which we calculate in several steps.  We start by calculating the
spectral density~$\rho(z)$ of the matrix~$\mat{X}$ alone, whose average
value in the random ensemble can be written in terms of the imaginary part
of the Stieltjes transform:
\begin{equation}
\rho(z) = -{1\over\pi} \Im \av{\Tr\,(z\mat{I}-\mat{X})^{-1}},
\label{eq:rho}
\end{equation}
where $\av{\ldots}$ indicates the ensemble average.  The average of the
trace can be expanded in powers of~$\mat{X}$ as
\begin{equation}
\av{\Tr(z\mat{I}-\mat{X})^{-1}}
  = {1\over z} \sum_{k=0}^\infty {\Tr\av{\mat{X}^k}\over z^k},
\label{eq:stieltjes}
\end{equation}
where the individual terms take the form
\begin{equation}
\Tr\av{\mat{X}^k}
    = \sum_{i_1\ldots i_k} \av{X_{i_1i_2} X_{i_2i_3} \ldots X_{i_ki_1}}.
\end{equation}
Since the elements of~$\mat{X}$ have mean zero, any term in this sum that
contains any variable just once will average to zero.  Moreover, terms
containing any variable more than twice become negligible when the average
degree of the network is much greater than one, so that the only terms
remaining are those for which $k$ is even and which contain each variable
exactly twice.  Geometrically, the sequence of indices in these terms takes
the form of an Euler tour of a rooted plane tree, with a factor of
$\av{X_{ij}^2}$ on each edge, whose average value is
$\half(\pin+\pout)=(\cin+\cout)/2n$.  Writing $k=2m$ with $m$ integer,
there are $n^{m+1}$ ways to choose the $m+1$ vertices of the tree and the
number of topologically distinct rooted plane trees with this many vertices
is equal to the Catalan number~$C_m$.  Thus
\begin{align}
\Tr\av{\mat{X}^{2m}} &= n^{m+1} \biggl( {\cin+\cout\over 2n} \biggr)^m C_m
                  \nonumber\\
               &= n \bigl[ \half(\cin+\cout) \bigr]^m C_m.
\label{eq:trx}
\end{align}
Combining this result with Eq.~\eqref{eq:stieltjes}, we have
\begin{align}
\av{\Tr(z\mat{I}-\mat{X})^{-1}}
  &= {n\over z} \sum_{m=0}^\infty \bigl[ \half(\cin+\cout) \bigr]^m
     C_m/z^{2m} \nonumber\\
  &= {n\over\cin+\cout}\,\Bigl[z - \sqrt{z^2-2(\cin+\cout)}\Bigr].
\label{eq:trace}
\end{align}
Then the spectral density, Eq.~\eqref{eq:rho}, is
\begin{equation}
\rho(z) = (n/\pi) {\sqrt{2(\cin+\cout)-z^2}\over\cin+\cout},
\label{eq:semicircle}
\end{equation}
which is a modified form of the classic Wigner semicircle law for random
matrices.  Note that the density of eigenvalues increases with~$n$, which
implies that the fluctuations in the values vanish as~$n\to\infty$.

Armed with this result we can now calculate the spectrum of the adjacency
matrix $\mat{A} = \av{\mat{A}} + \mat{X}$, but again we take the
calculation in stages, starting with the simpler exercise of calculating
the spectrum of the matrix
\begin{equation}
\mat{B} = \half (\cin-\cout)\,\vec{u}\vec{u}^T + \mat{X}
        = \mat{A} - \half (\cin+\cout)\,\vec{1}\vec{1}^T.
\label{eq:modmatrix}
\end{equation}
Note that $\half (\cin+\cout)\,\vec{1}\vec{1}^T$ is the uniform matrix with
all elements equal to $\half(\pin+\pout)$, which is the average
probability~$p$ of an edge in the entire network.  Hence the elements of
$\mat{B}$ are $B_{ij} = A_{ij} - p$.  This matrix is of interest in its own
right.  It is the so-called modularity matrix, which forms the basis for
the modularity maximization method of community detection.  The modularity
matrix is usually defined by $B_{ij}=A_{ij}-P_{ij}$ where $P_{ij}$ is the
expected value of the adjacency matrix element in a null model containing
no community structure.  The most commonly used null model is the
configuration model, a random graph with specified degree distribution, but
in the present case, for which all vertices have the same expected degree,
the null model is just a standard Erd\H{o}s--R\'enyi random graph with
$P_{ij}=p$ for all~$i,j$, leading to the definition in
Eq.~\eqref{eq:modmatrix}.  Thus our calculation will in this case give us
also the spectrum of the modularity matrix.

The general form of the matrix~$\mat{B}$ is that of a rank-1
matrix~$\vec{u}\vec{u}^T$ plus a random perturbation, a form that has been
studied in the mathematical literature.  Following an argument
of~\cite{BN11,CDF09}, let $z$ be an eigenvalue of this matrix and $\vec{v}$
be the corresponding normalized eigenvector, so that
\begin{equation}
\bigl[ \half(\cin-\cout) \,\vec{u}\vec{u}^T + \mat{X} \bigr] \vec{v}
  = z\vec{v}.
\end{equation}
A rearrangement gives $(z\mat{I}-\mat{X})\vec{v} = \half(\cin-\cout)
\,\vec{u}\vec{u}^T\vec{v}$, where $\mat{I}$ is the identity.  Multiplying
by $\vec{u}^T(z\mat{I}-\mat{X})^{-1}$ and cancelling a factor of
$\vec{u}^T\vec{v}$, we find that
\begin{equation}
{2\over\cin-\cout} = \vec{u}^T (z\mat{I}-\mat{X})^{-1} \vec{u}
  = \sum_{i=1}^n {(\vec{u}^T\vec{x}_i)^2\over z-\lambda_i},
\label{eq:perturb1}
\end{equation}
where $\lambda_i$ is the $i$th eigenvalue of~$\mat{X}$ and~$\vec{x}_i$ is
the corresponding eigenvector.

\begin{figure}
\begin{center}
\includegraphics[width=8cm]{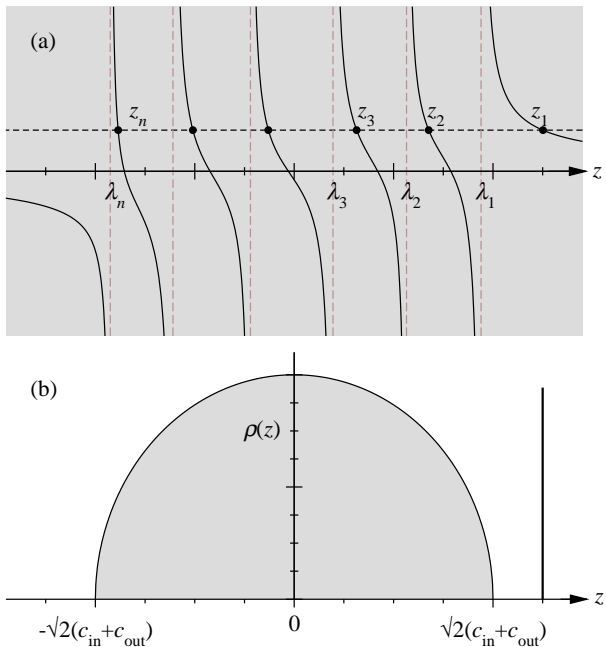}
\end{center}
\caption{(a)~The solid curve represents the right-hand side of
  Eq.~\eqref{eq:perturb1} while the dashed horizontal line represents the
  left-hand side.  The points at which the two cross, indicated by the
  dots, are the solutions~$z_i$ of the equation, which necessarily fall
  between the eigenvalues~$\lambda_i$ of the matrix~$\mat{X}$ (vertical
  dashed lines).  (b)~The spectrum of the modularity matrix is the same of
  that of the random matrix~$\mat{X}$ (semicircle), except for the highest
  eigenvalue~$z_1$, which is separate from the rest of the spectrum above
  the transition point given in Eq.~\eqref{eq:phase}.}
\label{fig:spectrum}
\end{figure}

The solutions of this equation, which give the eigenvalues~$z$ of the
modularity matrix, are represented graphically in Fig.~\ref{fig:spectrum}a.
The right-hand side of the equation has poles at $z=\lambda_i$ for all~$i$
and, as the figure shows, this means that the eigenvalues must satisfy
$z_1\ge\lambda_1\ge z_2\ge\lambda_2\ge\ldots\ge z_n\ge\lambda_n$, where
both sets of eigenvalues are numbered in order from largest to smallest.
These inequalities place bounds on the eigenvalues~$z_2\ldots z_n$ that
become tight as $n\to\infty$, meaning that the spectrum of the modularity
matrix is asymptotically identical to that of the random matrix~$\mat{X}$.

The only exception is the highest eigenvalue~$z_1$, which is bounded below
by~$\lambda_1$ but unbounded above.  To calculate this eigenvalue we note
that since $\mat{X}$ is a random matrix, its eigenvectors are also random,
so that cross-terms cancel in the quantity~$(\vec{u}^T\vec{x}_i)^2$ and the
average value is simply $|\vec{x}_i|^2/n = 1/n$.  Taking the average
of~\eqref{eq:perturb1} over the random matrix ensemble in the limit of
large~$n$ then gives
\begin{align}
{2\over\cin-\cout} &= {1\over n} \av{\sum_{i=1}^n {1\over z-\lambda_i}}
   = {1\over n} \av{\Tr(z\mat{I}-\mat{X})^{-1}} \nonumber\\
  &= {z - \sqrt{z^2-2(\cin+\cout)}\over\cin+\cout},
\label{eq:perturb2}
\end{align}
where we have used Eq.~\eqref{eq:trace}.  Rearranging for~$z$, we get our
expression for the leading eigenvalue~$z_1$:
\begin{equation}
z_1 = \half(\cin-\cout) + {\cin+\cout\over\cin-\cout}.
\label{eq:z1}
\end{equation}

We can use the same method to deduce the spectrum of the full adjacency
matrix also.  From Eq.~\eqref{eq:modmatrix} we see that the adjacency
matrix takes the form $\mat{A} = \half(\cin+\cout)\,\vec{1}\vec{1}^T +
\mat{B}$, which is again a rank-1 matrix plus a random perturbation.  By
the same argument as before, we can show that this matrix has all
eigenvalues the same (to within tight bounds) as those of the modularity
matrix, except again for the leading eigenvalue, whose value can be
calculated from a relation of the form~\eqref{eq:perturb1}.  The end result
is that the lower $n-2$ eigenvalues of the adjacency matrix have the same
spectrum as the random matrix~$\mat{X}$ and the top two have the
values~$z_1$, Eq.~\eqref{eq:z1}, and
\begin{equation}
z_2 = \half(\cin+\cout) + 1.
\label{eq:z2}
\end{equation}
With this result, we now have the complete spectrum for both the adjacency
matrix and the modularity matrix.

Let us focus on the modularity matrix.  The spectrum is depicted in
Fig.~\ref{fig:spectrum}b and consists of the continuous semicirclar band of
eigenvalues, Eq.~\eqref{eq:semicircle}, plus the single eigenvalue~$z_1$,
Eq.~\eqref{eq:z1}.  If the network contained no community structure, then
$z_1$ would not be separated from the continuous band as it is here.  So
long as it is well separated the spectrum shows clear evidence of the
existence of community structure and one can reasonably say that a
calculation of the spectrum constitutes positive ``detection'' of that
structure.  Moreover, the signs of the elements of the leading eigenvector
provide a good guide to the community division of the network, and indeed
this particular method for community identification can be derived directly
as a spectral version of the standard method of modularity
maximization~\cite{Newman06b}.  If, however, the position of the leading
eigenvalue passes the edge of the continuous band, the spectrum no longer
shows evidence of community structure and spectral algorithms based on the
corresponding eigenvector will fail.  One might imagine that this point
would arrive when $\cin=\cout$, which is the point at which the network
contains no community structure at all, but this is not the case.  From
Eq.~\eqref{eq:semicircle} we see that the end of the continuous band falls
at $z = \sqrt{2(\cin+\cout)}$ and, setting $z_1$ from Eq.~\eqref{eq:z1}
equal to this value, we find that we lose the ability to detect community
structure at an earlier point, when
\begin{equation}
\cin - \cout = \sqrt{2(\cin+\cout)}.
\label{eq:phase}
\end{equation}
This value sets a detectability threshold beyond which the communities are
present but cannot be detected.  For $\cin-\cout$ smaller than this value,
but greater than zero, community structure is present in the network in the
sense that the average probability of edges within groups is measurably
higher than that between groups, but we nonetheless fail to find the
communities using our spectral method.  One can generalize the calculation
to networks with a larger number~$q$ of communities and we find that a
similar transition happens at the point
\begin{equation}
\cin - \cout = \sqrt{q[\cin+(q-1)\cout]}.
\label{eq:transition}
\end{equation}
The existence of a transition of this kind, though not its precise
location, was demonstrated previously using different methods by Reichardt
and Leone~\cite{RL08} and there are also close connections between our
calculation and the theory of disordered systems~\cite{KTJ76}.

One might imagine this transition to be a particular property of the
spectral method we have considered.  Perhaps a different modularity
maximization algorithm, one not based on spectral techniques, or a
different type of community detection method altogether, would be able to
get past this detectability threshold.  This, however, is also not the
case.

In recent work, Decelle~\etal~\cite{DKMZ11a} have used arguments based on
the cavity method of statistical physics to demonstrate the existence of a
transition akin to the one above in another community detection method, a
Bayesian maximum-likelihood method based on directly fitting the stochastic
block model to a network.  Moreover, their transition falls at the same
position as that of Eq.~\eqref{eq:transition}.  The importance of this
result stems from the fact that if we know the model from which a network
is drawn, then fitting directly to that model is provably the optimal way
of recovering the parameters of the model used to generate the
network---including, in this case, the community structure.  Thus, as
Decelle~\etal\ have pointed out, their maximum-likelihood method is an
optimal method in the sense that no method can detect communities in the
regime where their method fails.  Unfortunately fitting to the stochastic
block model turns out to be a poor method of community detection for
real-world networks~\cite{KN11a,DKMZ11a}, but its optimality in the present
case is a useful result nonetheless.  It implies, given that the
detectability transition falls in the same place as for the spectral
modularity method, that the modularity method is also optimal in the same
sense: no other method will detect communities in the network when the
modularity method does not~\cite{note}.

We can take these calculations further.  For instance, we can calculate the
expected fraction of vertices classified correctly by the spectral
algorithm.  We can show that the elements of the leading
eigenvector~$\vec{v}$ of the modularity matrix are equal
to~$\pm\alpha/\sqrt{n}$ plus Gaussian perturbations with variance
$(1-\alpha^2)/n$, where
\begin{equation}
\alpha^2 = {(\cin-\cout)^2-2(\cin+\cout)\over(\cin-\cout)^2}.
\end{equation}
Then the fraction of elements that retain the correct sign and hence give
correct classifications of the corresponding vertices is
$\half[1+\erf\sqrt{\alpha^2/2(1-\alpha^2)}]$, where $\erf x$ is the
Gaussian error function.

\begin{figure}
\begin{center}
\includegraphics[width=\columnwidth]{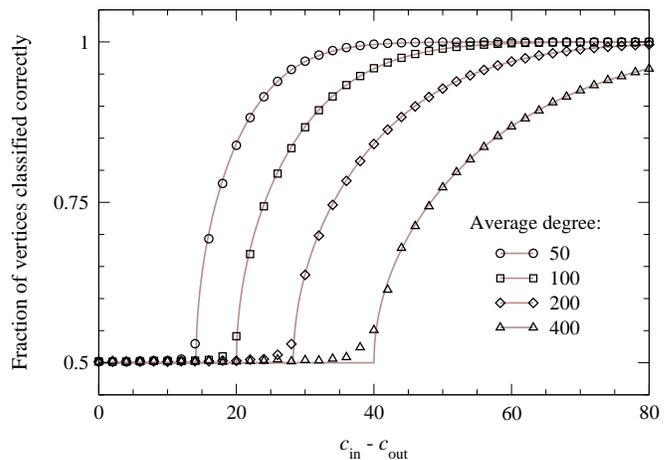}
\end{center}
\caption{The fraction of vertices correctly classified by the spectral
  modularity algorithm in networks generated using the block model studied
  here, as a function of $\cin-\cout$, for four different values of the
  average degree as indicated.  Points are numerical measurements for
  networks of $100\,000$ vertices, averaged over 25 networks each; solid
  curves represent the analytic prediction.  The phase transition at which
  the algorithm fails is clearly visible in each curve.}
\label{fig:results}
\end{figure}

Figure~\ref{fig:results} shows a plot of this quantity as a function of
$\cin-\cout$ for networks with several different values of the average
degree, along with results for the same quantity from actual applications
of the spectral modularity algorithm to networks generated using the
stochastic block model.  As the figure shows, the agreement between the two
is excellent, except in the immediate vicinity of the phase transition,
where finite-size effects produce some rounding of the threshold.  The
fraction of correctly classified vertices (minus~$\frac12$) plays the role
of an order parameter for the detectability transition.  Since it is
continuous at the transition point, we have a continuous phase transition.

The calculations presented here could be extended in a number of additional
directions.  For instance, the results given are accurate for networks with
large average degree but for networks with smaller degree there are
additional corrections that corresponding to additional terms in the trace,
Eq.~\eqref{eq:trx}.  A calculation of these sub-leading terms would help to
complete the picture for low-degree networks.  Also our calculations all
use the standard stochastic block model, and although this is the model
most widely used for benchmark calculations and synthetic tests, other
models have been proposed, such as the degree-corrected block
model~\cite{KN11a} or more exotic models such as the LFR benchmark
networks~\cite{LFR08}.  It would be useful to know if results similar to
those described here can be derived for these more complex models.

The authors thank Cris Moore and Lenka Zdeborova for useful comments.  This
work was funded in part by the National Science Foundation under grants
CCF--1116115 and DMS--1107796 and by the Office of Naval Research under
grant N00014--11--1--0660.


\begin{thebibliography}{10}
\bibitem{note} Strictly speaking, the results of Ref.~\cite{DKMZ11a} say
  only that no algorithm will succeed that works in polynomial time---it
  may be possible, at least in certain parameter regimes, for
  non-polynomial algorithms to detect the community structure.
  Non-polynomial algorithms, however, are prohibitively slow for all but
  the smallest of networks, making communities effectively undetectable
  past the transition point.
\end{thebibliography}
\end{document}